# Insulating phase of a two-dimensional electron gas in MgZnO/ZnO heterostructure below $\nu = 1/3$

Y. Kozuka,[1,*] A. Tsukazaki,[2,3] D. Maryenko,[4] J. Falson,[1] S. Akasaka,[5] K. Nakahara,[5] S. Nakamura,[6,7] S. Awaji,[7] K. Ueno,[4] and M. Kawasaki[1,2,4,8]

[1]*Institute for Materials Research, Tohoku University, Sendai 980-8577, Japan*

[2]*Department of Applied Physics and Quantum-Phase Electronics Center (QPEC), University of Tokyo, Tokyo 113-8656, Japan*

[3]*PRESTO, Japan Science and Technology Agency (JST), Tokyo 102-0075, Japan*

[4]*WPI Advanced Institute for Materials Research, Tohoku University, Sendai 980-8577, Japan*

[5]*Interdisciplinary Devices R&D Center, ROHM Co., Ltd., Kyoto 615-8585, Japan*

[6]*Center for Low Temperature Science, Tohoku University, Sendai 980-8577, Japan*

[7] *High Field Laboratory for Superconducting Materials, Institute for Materials Research, Tohoku University, Sendai 980-8577, Japan*

[8]*CREST, Japan Science and Technology Agency (JST), Tokyo 102-0075, Japan*

## Abstract

We report magnetotransport properties of a two-dimensional electron gas confined at MgZnO/ZnO heterointerface in a high magnetic field up to 26 T. High electron mobility and low charge carrier density enabled the observation of the fractional quantum Hall state $\nu = 1/3$. For an even lower charge carrier density, we observe a transition from quantum Hall liquid to an insulator below the filling factor 1/3. Because of the large electron effective mass in ZnO, we suggest the MgZnO/ZnO heterostructure to be a prototype system for highly correlated quantum Hall physics.

[*]kozuka@ap.t.u-tokyo.ac.jp

Two-dimensional (2D) charge carriers with high mobility can possess a variety of quantum phases such as integer and fractional quantum Hall states.[1-3] With increasing mobility, 2D carriers additionally exhibit intriguing phenomena: even denominator fractional states,[4] 2D metal-insulator transition,[5] and microwave-induced zero-resistance states.[6] Many of these studies have been carried out in a few limited material systems, mostly in GaAs, because extremely clean 2D charge carrier systems have been obtained.[7,8] Recently, however, novel 2D material systems have shown unprecedented findings, which are absent in conventional semiconductors, and disclosed new aspects of high-mobility 2D carriers.[9-14] For instance, the electrons in graphene behave like massless charge carriers owing to their linear dispersion,[9,10] which leads to a suggestion to observe relativistic effects in solids.

Oxide heterostructures are another example of a material system in which new phenomena may emerge. For example, high-mobility 2D electrons were successfully formed in $SrTiO_3$ heterostructures. Not only the well-known 2D Shubnikov-de Haas oscillations but also the transition to 2D superconducting state was demonstrated.[12,13] The MgZnO/ZnO heterointerface is another example of an oxide heterostructure hosting a 2D electron gas (2DEG). The large electron effective mass of ZnO is expected to result in strong electron-electron correlation effects represented by the interaction parameter $r_s = (n\pi)^{-1/2} m^* e^2 / 4\pi\hbar^2 \varepsilon$, where $n$ is the charge carrier density, $m^*$ is the effective mass, $e$ is the elementary electric charge, $\hbar$ is the Planck's constant divided by $2\pi$, and $\varepsilon$ is the dielectric constant. The recent observation of fractional quantum Hall effect in samples with mobility of 180,000 $cm^2$ $V^{-1}$ $s^{-1}$ and $r_s \sim 10$ demonstrates this concept.[14] Large $r_s$ parameters had been also obtained in conventional quantum Hall systems. In AlAs heterostructures, the interaction parameter reached a value of 10,[15] while the largest reported parameter $r_s$ = 36 was obtained for 2D hole gases in GaAs.[16] In all these materials, the carrier effective mass is large, which explains the large interaction parameter. However, these materials have multiple anisotropic Fermi surfaces, which may complicate the understanding of various phenomena. In contrast, the Fermi surface in ZnO is a sphere and thus similar to 2DEG in GaAs. Therefore, ZnO may be an ideal system to investigate quantum Hall physics with strong electron correlation. Here, we studied low-temperature magnetotransport of a 2DEG in MgZnO/ZnO heterostructure with improved electron mobility of 300,000 cm$^2$ V$^{-1}$ s$^{-1}$ in a magnetic field up to 26 T. Compared to previous studies,[14] the charge carrier density was reduced, which made it feasible to observe the fractional quantum state $\nu$ = 1/3 in oxide material for the first time. In addition, we investigated high-field magnetotransport in a sample with diluted charge carriers at filling factors below $\nu$ = 1/3. As a result, we found a magnetic-field-induced insulating phase, which is either disorder-induced quantum Hall insulator (QHI) or Wigner solid (WS) induced by electron correlation as observed in the 2D GaAs systems.[17-21]

The details of sample preparation were explained in Ref. 14. Briefly, we grew a 700 nm ZnO buffer layer followed by a 500 nm MgZnO layer on a Zn-polar ZnO(0001) single crystal substrate (Tokyo Denpa Co., Ltd.) by molecular beam epitaxy at 750 °C. Magnesium content of the MgZnO layer was 1.0 at.%, as measured by secondary ion mass spectroscopy. As a key ingredient of the mobility enhancement, we employed a pure ozone generator (Meidensya Co.) as the oxygen source. A Hall bar structure was etched using ion milling followed by Ti evaporation to form ohmic contacts. On top of the structure, an $Al_2O_3$ gate insulator was deposited by atomic layer deposition. Subsequently, Ni/Au gate electrode was evaporated on top of the Hall bar to enable tuning of the carrier density. The transport experiments were performed in a dilution refrigerator only at a base temperature of 40 mK due to unreliable control for higher temperatures in the present experimental setup. The sample, immersed in $^3$He-$^4$He mixture, was placed in a perpendicular magnetic field created by a cryogen-free hybrid magnet composed of an outer 8.5 T superconducting magnet and an inner 19 T water-cooled resistive magnet at the High Field Laboratory for Superconducting Materials, IMR, Tohoku University. While the polarity of 19 T resistive magnet can be reversed, the field direction of the superconducting magnet is always fixed. The sample was excited with a sinusoidal current of 10 nA with a frequency of 11 Hz. The longitudinal voltage $R_{xx}$ and Hall voltage $R_{xy}$ were measured simultaneously using the lock-in technique.

We first measured the basic transport characteristics of our 2DEG. Figure 1(a) shows that the electron mobility ($\mu$) increases with decreasing temperature, and reaches a maximum mobility of 300,000 cm$^2$ V$^{-1}$ s$^{-1}$ at 40 mK. The dependence of $\mu$ on charge carrier density at 40 mK is depicted in Fig. 1(b), where carrier density was varied by applying the voltage ($V_G$) to the top gate [inset of Fig. 1(b)]. The electron mobility reduces drastically below $n \sim 7 \times 10^{10}$ cm$^{-2}$ but saturates at $n = 2.0 \times 10^{11}$ cm$^{-2}$ when a gate voltage of 1.0 V is applied.

Given these basic transport properties, we then measured $R_{xx}$ and $R_{xy}$ in the magnetic field for a carrier density of $2.0 \times 10^{11}$ cm$^{-2}$. Relative to zero magnetic field the magnetotransport traces do not show significant asymmetry between -19T and +19T, which indicates no mixing between $R_{xx}$ and $R_{xy}$. Figures 2(a) and 2(b) show the traces of $R_{xx}$ and $R_{xy}$ in a magnetic field up to 26 T. $R_{xx}$ showed clear Shubnikov-de Haas oscillations starting at a low magnetic field ($B$) of 0.3 T, which suggests high sample quality. As in the previous report,[14] we observed the fractional states at $\nu = 5/3$ and $4/3$. Around filling factor $\nu = 1/2$ in Fig. 2(b), a number of fractional states $\nu = 2/3$, $3/5$, and $2/5$ are observed. The most remarkable observation is the appearance of the fundamental fractional state $\nu = 1/3$ at $B = 25.5$ T. We note that this system is the fourth material to exhibit this fractional state, following GaAs/AlGaAs,[2] Si/SiGe,[22] and graphene.[23,24] In addition to these fractional states, we observed a sharp peak in $R_{xx}$ between $\nu = 2/5$ and $\nu = 1/3$, which has been attributed to WS in GaAs,[18-21] although detailed studies are necessary to reveal the origin in our system. Another feature of $R_{xx}$ is the distinct behavior at $\nu = 3/2$ and $\nu = 1/2$. While $R_{xx}$ has an indication of a minimum at $\nu = 3/2$, $R_{xx}$ shows a negative magnetoresistance at $\nu = 1/2$. According to Kalmeyer *et al.*, 2DEG exhibits a metallic and an insulating phases at corresponding filling factors, which is interpreted as a stronger electron localization at $\nu = 1/2$.[25]

To increase the correlation parameter $r_s$ and thus to enhance the electron correlation effect, we reduced the electron density by applying the top gate voltage [see Fig. 1(b)]. This allowed us also to access filling factors below $\nu = 1/3$. Figure 2(c) shows $R_{xx}$ and $R_{xy}$ as a function of magnetic field with a reduced carrier density of $1.6 \times 10^{11}$ cm$^{-2}$. The integer quantum Hall states remain stable as is evident at $\nu = 1$, while fractional states become less pronounced although a dip in $R_{xx}$ and quantized $R_{xy}$ are still distinguished at $\nu = 1/3$. This nonzero $R_{xx}$ could be the effect of lateral inhomogeneity as is also seen at $\nu = 1$ in Fig. 2(a). However, this does not affect the following discussions. Below $\nu = 1/3$, $R_{xx}$ shows a sharp increase, while $R_{xy}$ does not strongly deviate from its classical value ($B/ne$) as indicated by the dashed line in Fig. 2(c). Such insulating behavior has been attributed to the formation of either QHI or WS.[17-21] In order to further investigate this insulating behavior in the low $\nu$ limit, we measured $R_{xx}$ and $R_{xy}$ as a function of charge carrier density at a fixed magnetic field of 20 T as shown in Fig. 3(a). According to the behavior of $R_{xx}$ and $R_{xy}$, $\nu$ values are classified mainly into three regions. Region A ($\nu > 0.31$), where the quantum Hall effect is observed as described above. Region B ($0.26 < \nu < 0.31$), where $R_{xx}$ diverges and reaches 40 k$\Omega$, while $R_{xy}$ follows the classical value indicated by the dashed line in Fig. 3(a). Such behavior is characteristic for the formation of QHI or WS. And finally, region C ($\nu < 0.26$), where both $R_{xx}$ and $R_{xy}$ diverge and the sample becomes insulating. The separation between the regions is indistinct and therefore the borders between the regions are indicated with a light gray bar in Fig. 3(a).

To characterize the insulating phase of the 2DEG, the temperature dependence of $R_{xx}$ or alternatively the differential resistance d$V$/d$I$ is a possible way.[26,27] Because of our experimental limitation, here we measured four-terminal differential resistance d$V$/d$I$ at several filling factors to assert the conducting phase of our sample. The insulating phase, e.g. QHI or WS, is characterized by a negative d$^2V$/d$I^2$, while positive d$^2V$/d$I^2$ reflects quantum Hall liquid. Figure 3(b) shows a map of d$V$/d$I$ as functions of $I_{dc}$ and $\nu$. This figure visually represents a transition from quantum Hall liquid to the insulating phase around $\nu = 0.28$ as displayed in Figs. 3(c) – 3(f). Near $\nu = 1/3$, i.e. region A, d$V$/d$I$ increases with increasing $I_{dc}$ as shown in Fig. 3(c) and confirms the quantum Hall state. With decreasing the charge carrier density to region B, $R_{xx}$ shows insulating behavior while $R_{xy}$ remains classical. This is also reflected in the change of the $I$-$V$ characteristics.

Close to the middle of region B, i.e. at $\nu = 0.28$, the differential resistance starts developing a characteristic shape for an insulator, but the admixture of the metallic phase is still visible. This behavior indicates that this filling factor is right at the transition between quantum Hall liquid to an insulating state. As the charge carrier density is slightly reduced, $\nu = 0.26$, the differential resistance shows a pronounced insulating behavior. This filling factor falls into the border between region B and region C. Here, $R_{xx}$ diverges and $R_{xy}$ does not deviate very much from the classical value. Thus, at this charge carrier density, QHI or WS is likely formed. At an even lower charge carrier density, e.g. region C, the *I-V* characteristic shows the insulating state of the sample [Fig. 3(f)]. This and the divergence of both $R_{xx}$ and $R_{xy}$ in region C shown in Fig. 3(a) assert a regular insulating state of the system, where charge carriers are localized.

High field insulating phase due to the formation of WS or QHI has been also observed in *n*-GaAs/AlGaAs[20] below $\nu = 1/5$ and both in *p*-GaAs/AlGaAs[21] and *n*-Si/SiGe[22] below $\nu = 1/3$. As a criterion, Tanatar *et al*. calculated that, in two dimensions, the critical carrier density of WS formation at zero magnetic field was determined as $r_s \approx 37$ (Ref. 28). With applying magnetic field, the critical $r_s$ is expected to decrease to 22 at $\nu = 1/3$ and reach zero at $\nu = 1/6.5$ (Ref. 29). In the current study, $r_s$ ranges from 8.5 to 11, using the parameters of ZnO ($m^* = 0.29 m_0$, $\varepsilon = 8.3 \varepsilon_0$; $m_0$ is the bare electron mass and $\varepsilon_0$ is the vacuum permittivity),[30] which is far from the expected critical $r_s$ value for WS at $\nu = 1/3$. In spite of the discrepancy of the critical $r_s$ between calculation and our results, the possibility of WS is not completely excluded due to the following argument. Landau level splitting is comparable to Coulomb energy (~ 10 meV) even at 26 T with $n = 1.0 \times 10^{11}$ cm$^{-2}$ in ZnO because of its large $m^*$. Therefore, we expect nontrivial Landau level mixing, which modifies formation energies of both fractional quantum Hall states and WS, leading to a lower criterion of $r_s$ value, as discussed in 2D hole gas of GaAs.[21] Since both QHI and WS show similar direct-current electrical behavior, radio-frequency spectroscopy studies are necessary by detecting a resonance mode of WS,[18,19] while QHI does not show such resonance because it is a single-particle localization.

In summary, we have studied magnetotransport properties of a high-mobility MgZnO/ZnO 2DEG up to 26 T. Owing to the improvement of the sample quality, we observed a number of fractional states around $\nu = 1/2$, including $\nu = 1/3$ for the first time in this system. By reducing electron density, we investigated further low filling factors below $\nu = 1/3$, resulting in a sharp increase of $R_{xx}$. Measurements of four-terminal differential resistance revealed that it originated from a transition from the $\nu = 1/3$ quantum Hall state to a high-field insulating state. As in previous reports for GaAs and Si/SiGe systems, this behavior can be attributed to the formation of quantum Hall insulator or Wigner solid. Despite these similarities to other semiconductors, we suggest that the large effective mass and the simple band structure of ZnO make this material an ideal system to study correlation effects in quantum Hall physics. Further studies will unveil the potential of this system for unexplored quantum phases of two-dimensional carriers.

This work was partly supported by Grant-in-Aids for Scientific Research No. 22840004 (Y.K.), No. 21686002, and No. 22103004 (K.U.) from MEXT, Japan, and by the Japan Society for the Promotion of Science (JSPS) through its "Funding Program for World-Leading Innovative R&D on Science and Technology (FIRST Program)" (A.T.) as

**Figure Captions:**

**FIG. 1:** (a) Temperature dependence of electron mobility estimated from low-field Hall effect with a carrier density of $2.0 \times 10^{11}$ cm$^{-2}$. (b) Electron mobility with varying carrier density by electrostatic gating at 40 mK. The inset shows carrier density as a function of gate voltage. The dotted line is a result of fitting by a linear function.

**FIG. 2:** (color online) $R_{xx}$ and $R_{xy}$ (a) from $B = 0$ T to 11 T and (b) from 9 T to 26 T with a carrier density of $2.0 \times 10^{11}$ cm$^{-2}$, measured at 40 mK. (c) $R_{xx}$ and $R_{xy}$ with a carrier density of $1.6 \times 10^{11}$ cm$^{-2}$. The dashed line shows classical Hall resistance extrapolated from low field. Magnified $R_{xx}$ curve is also shown.

**FIG. 3:** (color online) (a) $R_{xx}$ and $R_{xy}$ as a function of $\nu$ [bottom axis of (b)] with reducing $n$ (top axis) at 40 mK under a magnetic field of 20 T, which are classified into three regions separated by the borders, as discussed in the main text. Dashed curve shows $R_{xy}$ values expected from the low-field Hall effect. (b) Map of the differential resistance (d$V$/d$I$) as functions of $I_{dc}$ and $\nu$ in a logarithmic scale at a fixed magnetic field of 20 T. Inner white tics at the bottom axis indicate $\nu$ where $I$-$V$ characteristics are measured. Raw data of d$V$/d$I$ − $I_{dc}$ are shown at filling factors of (c) 0.33, (d) 0.28, (e) 0.26, and (f) 0.24.

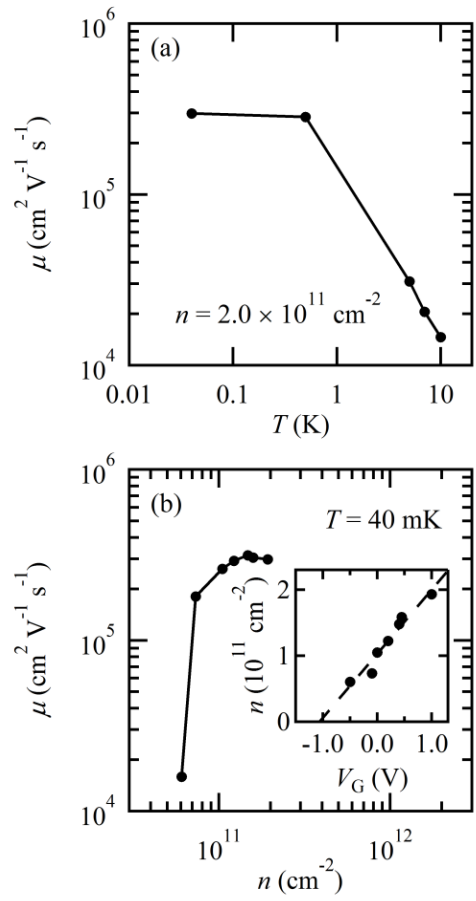

FIG. 1. Y. Kozuka et al.

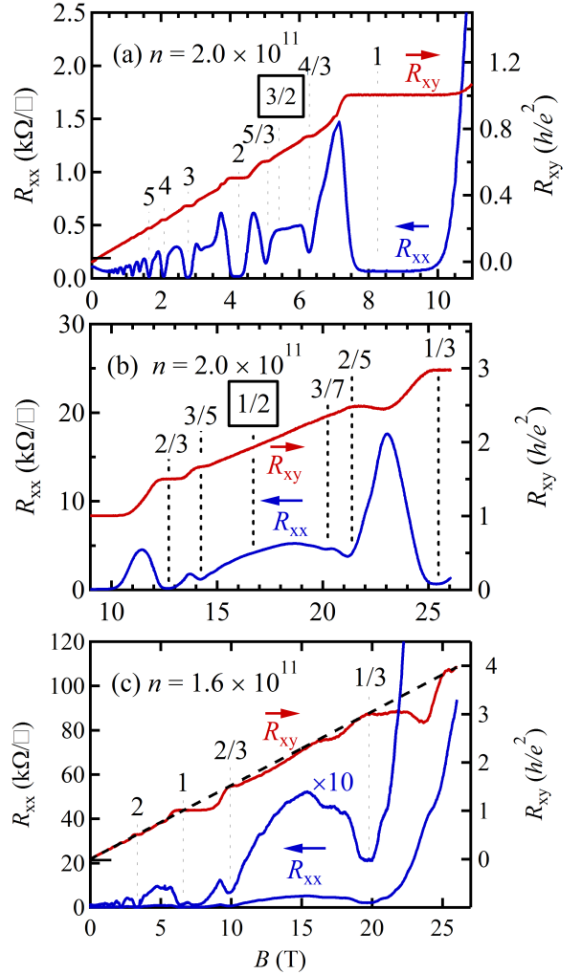

FIG. 2. Y. Kozuka et al.

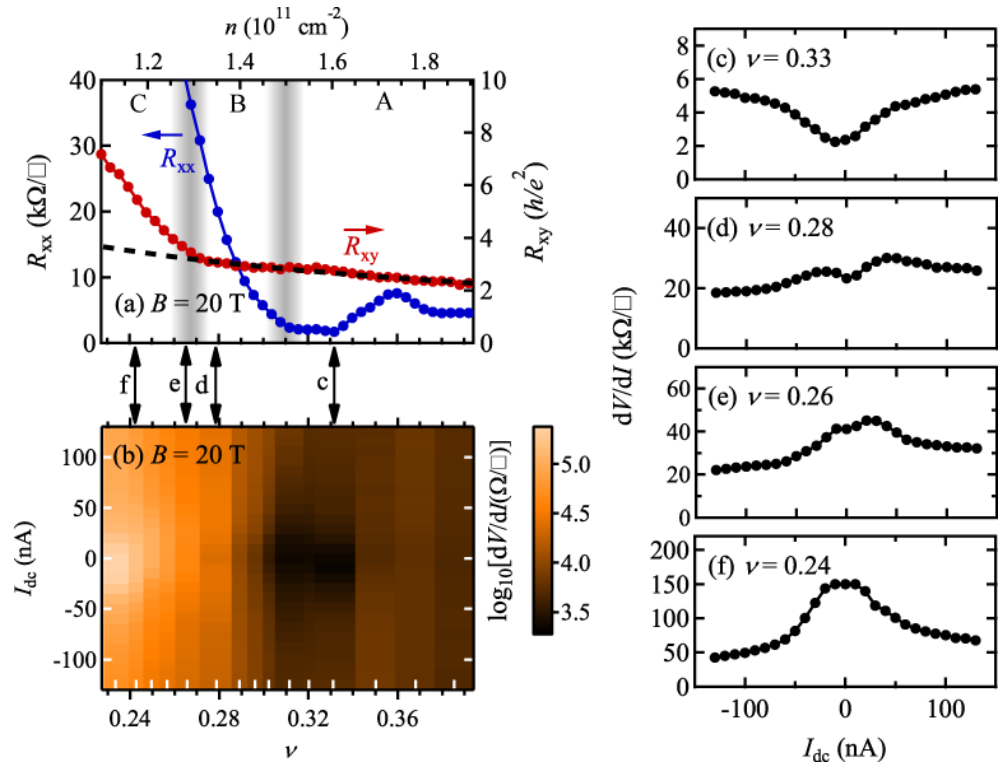

FIG. 3. Y. Kozuka et al.